\documentclass[10pt,twocolumn]{article}

\usepackage[T1]{fontenc}
\usepackage[utf8]{inputenc}
\usepackage{lmodern}
\usepackage{microtype}

\usepackage{amsmath,amssymb,amsfonts}
\usepackage{bm}

\usepackage{graphicx}
\usepackage{xcolor}
\usepackage{textcomp}

\usepackage{array}
\usepackage{multirow}
\usepackage{makecell}
\usepackage{tabularx}
\usepackage{booktabs}

\usepackage{algorithm}
\usepackage{algorithmic}

\usepackage{cite}

\usepackage[
  colorlinks=true,
  linkcolor=blue!45!black,
  citecolor=blue!45!black,
  urlcolor=blue!45!black
]{hyperref}

\usepackage[
  top=0.72in,
  bottom=0.72in,
  left=0.68in,
  right=0.68in,
  columnsep=0.24in
]{geometry}

\usepackage{titlesec}

\titleformat{\section}
  {\normalfont\large\bfseries}
  {\thesection.}
  {0.45em}
  {}

\titleformat{\subsection}
  {\normalfont\normalsize\bfseries}
  {\thesubsection}
  {0.45em}
  {}

\titleformat{\subsubsection}
  {\normalfont\normalsize\itshape}
  {\thesubsubsection}
  {0.45em}
  {}

\titlespacing*{\section}
  {0pt}{1.5ex plus .5ex minus .2ex}{0.8ex}

\titlespacing*{\subsection}
  {0pt}{1.2ex plus .4ex minus .2ex}{0.5ex}

\titlespacing*{\subsubsection}
  {0pt}{1ex plus .3ex minus .2ex}{0.3ex}

\graphicspath{{./images/}{../images/}{./}}

\renewcommand{\arraystretch}{1.15}

\title{%
  \textbf{SVM + CFAR + OR: A Hybrid Machine-Learning-CFAR Receiver
  for Low-False-Alarm PRACH Preamble Detection in 5G NR and Beyond}%
}

\author{%
  Filip Ferenc\thanks{Corresponding author:
  \href{mailto:263578@student.pwr.edu.pl}{263578@student.pwr.edu.pl}}
  \\
  \small Faculty of Information and Communication Technology,
  Wroc{\l}aw University of Science and Technology,
  Wroc{\l}aw, Poland
  \and
  Micha{\l} Szczachor
  \\
  \small Nokia, Wroc{\l}aw, Poland
}

\date{}

\newcommand{\keywords}[1]{%
  \par\noindent
  \textbf{Keywords:} #1
}

\begin{document}

\maketitle

\begin{abstract}
The physical random access channel (PRACH) is the initial access channel between the user equipment and the base station, and the ability to initiate transmission in a cellular network depends on its reliable operation. The growing device density in 5G and beyond-5G networks, massive machine-type communication, and the trend towards a machine-learning-driven radio interface make analytical preamble detectors increasingly limited in their ability to adapt to complex and dynamic radio environments, particularly at low signal-to-noise ratios. This article recasts preamble detection as a 64-class classification problem and introduces two model-agnostic contributions: a robust adaptive threshold that realizes the constant-false-alarm-rate (CFAR) principle directly in the domain of a classifier's decision scores (MAD-CFAR), and a hybrid receiver that fuses the learned detector with a classical CFAR detector on the power delay profile through a logical OR rule. Because both mechanisms are independent of the underlying classifier, the learned branch is instantiated with two models trained on the same fivefold-enlarged, GPU-generated dataset: a support vector machine (SVM) with principal component analysis, retained for its interpretability and analyzed with a nonlinear embedding and post-hoc attribution, and a lightweight one-dimensional convolutional neural network (CNN). On a held-out test set the SVM reaches 98.58\% accuracy (macro-averaged area under the curve, AUC, of 0.9991), whereas a lightweight CNN reaches 99.29\% accuracy (macro-AUC 1.0000) at substantially lower inference cost and memory footprint. The MAD-CFAR threshold drives the learned branch's false alarm rate to near zero on its noise calibration campaign, and the OR hybrid inherits the analytical detector's low, standard-compliant false alarm rate while widening the range of correct detection. Across six propagation channels and 258 conformance-style tests, the SVM-based hybrid gains several decibels of low-SNR sensitivity with no missed-detection regression relative to the analytical branch, a property that follows directly from the OR construction. Under the same false-alarm control the CNN is competitive in some channels and, at present, trails the SVM in high-delay-spread multipath, leaving room for further improvement.
\end{abstract}


\vspace{0.5em}

\noindent
\textit{This work has been submitted to the IEEE for possible publication. 
Copyright may be transferred without notice, after which this version may no longer be accessible.}

\section{Introduction}
The initial link establishment between user equipment (UE)
and base stations (BS) in 5G New Radio (NR) and emerging sixth-generation (6G)
networks relies on the random access (RA) procedure and the physical random
access channel (PRACH). Standardized preambles constructed from Zadoff--Chu (ZC)
sequences exhibit constant amplitude and zero cyclic autocorrelation, enabling
analytical correlation receivers to detect transmitted preambles and estimate
timing advance (TA).

However, target 6G network densities ($10^7$--$10^8$ devices/$\text{km}^2$)
under massive machine-type communication (mMTC) and AI-native radio architectures
expose critical limitations of classical correlation detectors. Under low
signal-to-noise ratios (SNR), severe multipath delay spreads, high Doppler
shifts, and heavy multiple-access interference, analytical Constant False Alarm
Rate (CFAR) detectors suffer from degraded sensitivity or excessive false
alarms. Excess false alarms waste uplink grants and deplete BS computing
resources, while missed detections cause connection setup failures.

To address these challenges, this article introduces \emph{SVM + CFAR + OR},
a novel hybrid PRACH receiver architecture, and establishes a broader,
model-agnostic \emph{ML + CFAR + OR} framework. The proposed solution combines
an analytical CFAR detector operating on the Power Delay Profile (PDP) with a
score-thresholded machine learning (ML) classifier via a logical OR decision
rule. The key methodological novelty is twofold: (i) an adaptive median
absolute deviation CFAR threshold (\emph{MAD-CFAR}) applied directly to raw
classifier decision scores, driving false alarms to zero on a 28-million
position noise campaign, and (ii) a logical OR decision fusion that guarantees
zero missed-detection regression relative to the analytical reference detector.
To prove model independence within the \emph{ML + CFAR + OR} framework, the
learned branch is instantiated with both a kernel Support Vector Machine (SVM
with PCA) and a lightweight 1-D Convolutional Neural Network (CNN), both
benchmarked on an identical 1.05-million sample dataset generated by a
commercial Nokia physical-layer simulator.

Prior ML PRACH literature (e.g., Modina \textit{et al.}~\cite{Modina2019},
Zehra \textit{et al.}~\cite{Zehra2022}, Guel \textit{et al.}~\cite{Guel2024},
Singh \textit{et al.}~\cite{Singh2024}) has focused primarily on binary
peak-validation ($k=2$) or timing assistance. In contrast, this work addresses
direct 64-class preamble identification from raw IQ samples under strict 3GPP
TS~38.141 base station conducted conformance requirements. Building upon our
earlier study~\cite{Ferenc_2025}, this paper extends the formulation through a
fivefold larger dataset, nonlinear t-SNE and SHAP explainability analyses,
score-domain MAD-CFAR thresholding, head-to-head SVM vs.\ CNN benchmarking,
and 258-scenario conformance validation.

The main contributions of this work are summarized as follows:
\begin{itemize}
  \item \emph{The SVM + CFAR + OR Receiver Architecture}: A novel hybrid PRACH
        receiver combining PDP-CFAR correlation and score-thresholded SVM
        classification via logical OR fusion, providing a mathematical guarantee
        of no missed-detection regression relative to the reference detector.
  \item \emph{Classifier-Agnostic ML + CFAR + OR Framework}: A robust
        score-domain adaptive thresholding mechanism (\emph{MAD-CFAR}) that
        enforces CFAR operating point control on raw decision scores, driving
        false alarms to zero regardless of the classifier architecture.
  \item \emph{Rigorous SVM vs. CNN Benchmarking}: A head-to-head evaluation
        of an interpretable OvA SVM with PCA and a lightweight 1-D CNN on the
        identical 1.05-million sample dataset, supported by t-SNE and
        SHAP/LIME explainability analyses.
  \item \emph{Commercial Conformance Validation}: Comprehensive evaluation with
        Nokia's commercial physical-layer simulator across six propagation
        channels and 258 conducted conformance scenarios under 3GPP TS 38.141-1.
\end{itemize}

The remainder of this article is organized as follows. Section~\ref{sec:background}
provides background on PRACH and analytical detection. Section~\ref{sec:method}
details the \emph{SVM + CFAR + OR} architecture and \emph{ML + CFAR + OR}
framework. Section~\ref{sec:setup} outlines the experimental setup,
Section~\ref{sec:results} presents the performance results and a comparison with existing approaches, and Section~\ref{sec:conclusion} concludes the paper.

\section{Background}\label{sec:background}
\subsection{PRACH and the Random Access Procedure}
During contention-based random access, the UE selects one of $64$ preambles
generated from ZC root sequences and cyclic shifts, as defined in 3GPP
TS~38.211~\cite{3GPP-TS-38.211}, and transmits it in a PRACH occasion. The BS
receiver must decide which preambles -- if any -- are present and estimate their
timing advance for the subsequent uplink grant. The constant-amplitude,
zero-autocorrelation property of ZC sequences ideally yields a sharp
correlation peak whose position encodes the TA.

\subsection{Analytical Detector}\label{subsec:analytical}
The reference analytical detector correlates the received signal with each ZC
root sequence, combines the per-antenna correlations into a PDP, and applies an
adaptive constant-false-alarm-rate (CFAR) threshold. The threshold is set from
a target false-alarm probability, the number of preambles, and the number of
receive antennas; the position of the first path in the detected window yields
the TA estimate. This detector offers a predictable, analytically controlled
FAR, but its sensitivity degrades when the modeling assumptions behind the CFAR
threshold break down -- for example, when the delay spread approaches the search
window length or when inter-carrier interference (ICI) grows at high UE speed.

\subsection{Learned Classifiers}
The SVM~\cite{Vapnik1995} is a maximum-margin classifier that, through the
kernel trick, constructs nonlinear decision boundaries in a high-dimensional
feature space. Multi-class problems are handled here with the One-vs-All (OvA)
scheme~\cite{Hsu2002}, yielding one decision score per preamble class. As an
alternative learned detector, a one-dimensional convolutional neural network
(CNN)~\cite{Goodfellow2016} applies shared, locally-connected filters along the
sample axis and, being far less expensive at inference, is well suited to a
real-time base-station receiver. In both cases it is the vector of raw,
per-class scores -- SVM decision values or CNN logits -- rather than hard labels,
that is thresholded by the mechanism introduced in
Section~\ref{subsec:madcfar}. This score-level interface is what makes the
proposed thresholding and fusion independent of the classifier.

\section{Proposed Method}\label{sec:method}
\subsection{Problem Formulation and Metrics}
Each PRACH occasion is mapped to a feature vector, and preamble detection is
posed as a 64-class classification problem with an additional
\emph{noise/no-preamble} outcome governed by adaptive score-domain thresholding.
Two metrics dictate receiver evaluation: the missed detection rate (MDR), i.e.,
the probability of failing to detect a transmitted preamble, and the false alarm
rate (FAR), i.e., the probability of declaring a preamble on a noise-only
position. Per 3GPP base station conducted conformance requirements
(TS~38.141-1~\cite{3GPP-TS-38.141-1}), an acceptable receiver must maintain
$P_\mathrm{FA}\le 10^{-3}$ while achieving a detection probability of at least
$99\%$.

\emph{Physical Modeling Scope.} In 3GPP NR architectures
(TS~38.211~\cite{3GPP-TS-38.211}), 64 cell preambles are generated via cyclic
shifts $N_\mathrm{CS}$ of Zadoff--Chu (ZC) root sequences. A UE transmission
is subject to a propagation delay (Timing Advance, TA) $d$, yielding
$r(n) = h(n) \circledast s_i(n - d) + w(n)$. Direct time-domain IQ feature
classification assumes coarse frame timing alignment or small-to-medium cell
configurations where propagation delay $d < N_\mathrm{CS}$. For large-cell
deployments ($d \ge N_\mathrm{CS}$), preamble identification is complemented by
PDP correlation or delay-invariant feature representations. Furthermore, while
single-preamble classification is evaluated as the core multi-class benchmark,
multi-label learning with binary cross-entropy is incorporated into the CNN
branch to support multi-preamble superposition in high-density access occasions.
Additionally, while 839-sample long preambles (Formats 0--3, 1.25/5~kHz SCS) are
evaluated as standardized macro-cell formats in 5G NR FR1, the proposed
\emph{ML + CFAR + OR} framework is format-agnostic and extends directly to 5G
NR short preamble formats ($L=139, 571$) by scaling the input feature length
$D=2L$.

\subsection{Dataset and Dimensionality Reduction}
The training dataset is generated with a professional LTE/5G physical-layer
simulator provided by Nokia, implemented in MATLAB and reproducing a full
LTE/5G network environment. Signals are recorded for all $64$ preambles across
$41$ SNR points from $-20$ to $+20$~dB in a topology of two base stations, each
equipped with two receive antennas. With a $10$~ms access-slot step over a
one-second simulation, each base station contributes $100$ occasions per SNR
point; treating the two base stations and their individual antenna paths as
independent observations yields about one million signals in total
($64\times41\times100\times2\times2\approx1.05\times10^{6}$). The classifier is
trained on four propagation
channels -- AWGN, EPA, EVA, and ETU -- whereas two further channels, TDLC300 and
TDLD30, are held out for testing only. Each complex preamble of $839$ samples is
decomposed into its real and imaginary parts, concatenated into a single feature
vector of length $D=1678$, and normalized before classification.

\begin{figure}[!t]
  \centering
  \includegraphics[width=0.86\columnwidth]{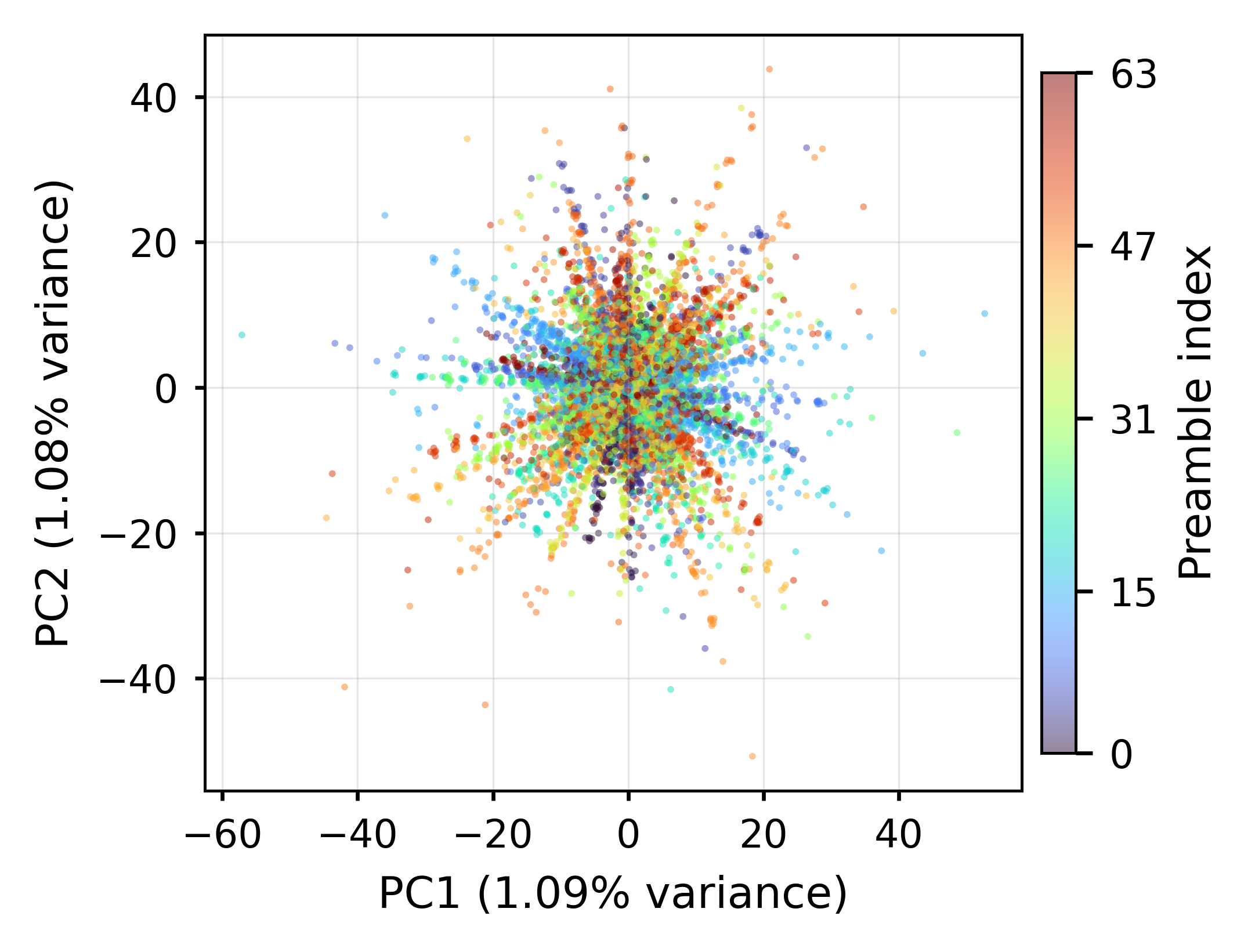}\\
  {\footnotesize (a) PCA projection onto the first two principal components.}\\[4pt]
  \includegraphics[width=0.86\columnwidth]{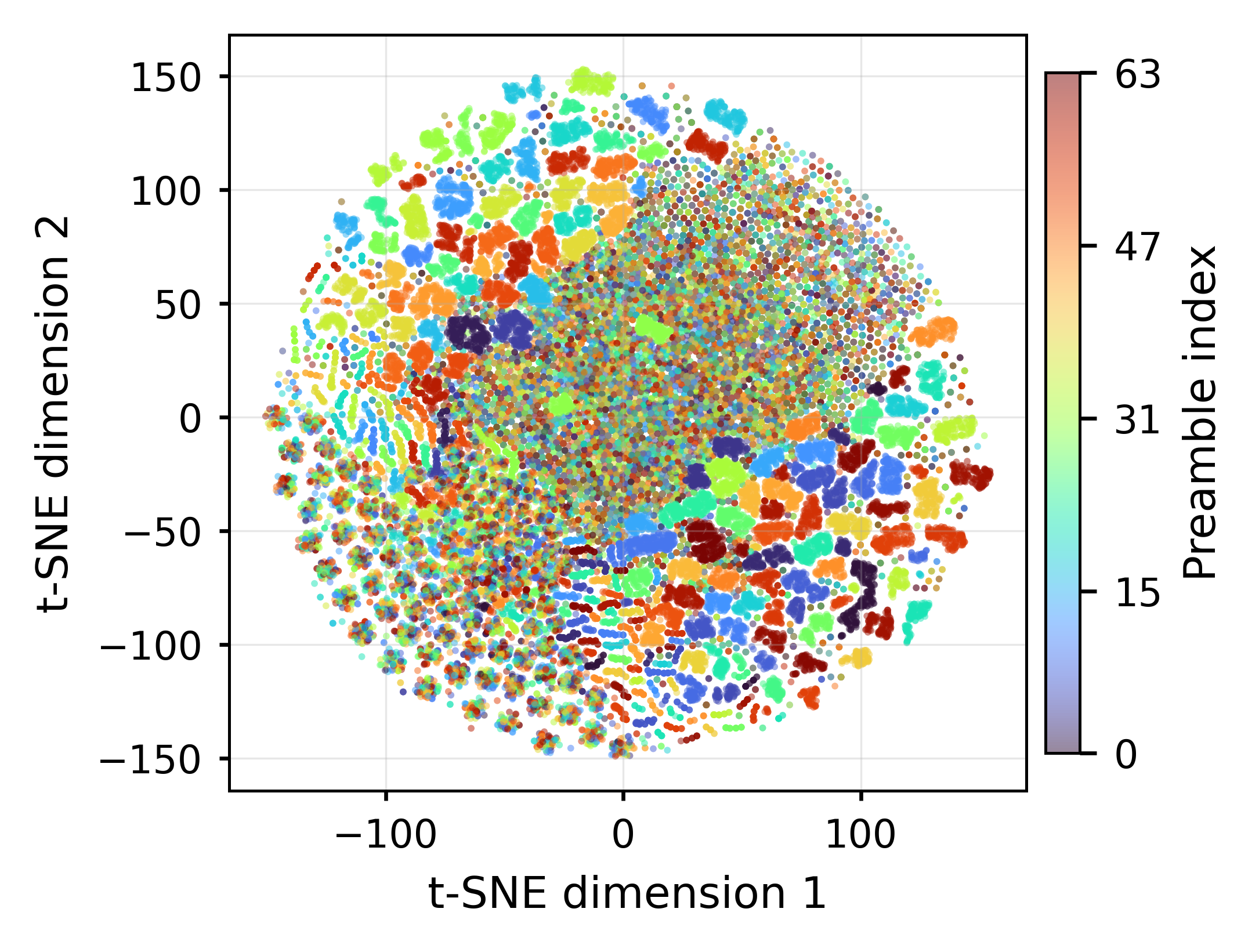}\\
  {\footnotesize (b) Nonlinear t-SNE embedding.}
  \caption{Two-dimensional embeddings of the PRACH preamble dataset, colored by
    preamble index. PCA (a) collapses the classes into a dense radial pattern,
    while t-SNE (b) shows only partial, condition-dependent separability: some
    preambles form compact clusters whereas others overlap heavily. As t-SNE
    preserves only local structure, cluster sizes and positions are not
    quantitatively meaningful.}
  \label{fig:embed}
\end{figure}

The raw $D=1678$-dimensional representation is high-dimensional; principal
component analysis (PCA)~\cite{brunton_svd_2019} is therefore applied as a
dimensionality-reduction stage, projecting each vector onto the $K=515$
principal components that retain $95\%$ of the variance (a reduction of over
$69\%$ from $D=1678$) before the classifier. Because the linear PCA projection alone does not
reveal the class structure of the data, the nonlinear t-SNE
method~\cite{vandermaaten2008tsne} is additionally used for visualization; as
t-SNE preserves only local neighborhood structure, cluster sizes and positions
are not interpreted quantitatively. Fig.~\ref{fig:embed} shows both embeddings:
PCA collapses all classes into a dense radial pattern (its first two components
explain only $2.17\%$ of the variance), whereas t-SNE reveals a strongly
condition-dependent structure in which compact, nearly single-class clusters
coexist with regions of heavy class overlap. This partial, nonlinear
separability motivates a classifier able to model complex decision boundaries,
i.e., the kernel SVM.

\subsection{Classifier Training: SVM and CNN}\label{subsec:training}
\emph{SVM Model Formulation.} The primary learned detector is instantiated as
a multi-class Support Vector Machine (SVM) utilizing a One-vs-All (OvA)
decomposition scheme across all $64$ standardized 3GPP PRACH preamble classes.
Given a PCA-reduced feature vector $\mathbf{x} \in \mathbb{R}^{K}$ (with
$K=515$), the $i$-th binary classifier outputs a raw decision score
$s_i(\mathbf{x})$ according to:
\begin{equation}
  s_i(\mathbf{x}) = \sum_{j \in \mathcal{SV}_i} \alpha_{i,j} y_{i,j} K(\mathbf{x}_j, \mathbf{x}) + b_i, \quad i \in \{1, \dots, 64\},
  \label{eq:svm_decision}
\end{equation}
where $\mathcal{SV}_i$ denotes the set of support vectors for class $i$,
$\alpha_{i,j} \ge 0$ are the dual Lagrange multipliers, $y_{i,j} \in \{-1, +1\}$
are the binary labels, and $b_i$ is the bias offset. A degree-2 polynomial kernel
is selected:
\begin{equation}
  K(\mathbf{x}_j, \mathbf{x}) = \left( \gamma \, \mathbf{x}_j^{\top} \mathbf{x} + c_0 \right)^2.
  \label{eq:poly_kernel}
\end{equation}
The degree-2 polynomial kernel is chosen a priori because it provides non-linear
boundary modeling while inducing lower support vector density and tighter
memory requirements than a Gaussian Radial Basis Function (RBF) kernel.
Hyperparameter optimization was conducted via a Bayesian tree-structured Parzen
estimator over $40$ trials, optimizing the soft-margin box constraint $C$,
kernel scale $\gamma$, and offset $c_0$, yielding optimal parameters
$C=0.480$, $\gamma=0.0303$, and $c_0=0.0297$. The entire training pipeline was
executed in a GPU-accelerated environment (RAPIDS cuML on an NVIDIA~H100 GPU)
on the WCSS computing cluster, enabling model fitting on a fivefold larger
dataset relative to earlier CPU-based implementations. Under five-fold
stratified cross-validation, the SVM classifier achieves $98.42\pm0.03\%$
accuracy, and on the held-out test set it reaches $98.58\%$ accuracy with a
macro-averaged AUC of $0.9991$ (all individual classes $>0.99$). For each
occasion, the model produces a raw decision score vector
$\mathbf{s} = [s_1, \dots, s_{64}]^{\top} \in \mathbb{R}^{64}$ that feeds the
score-domain thresholding stage.

\emph{CNN Architecture and Optimization.} To test the classifier-agnostic
property of the proposed framework, a lightweight 1-D Convolutional Neural
Network (CNN) is trained as an alternative learned detector on the \emph{exact
same dataset}. The CNN operates directly on raw normalized IQ signals
represented as two input channels ($2 \times 839$) rather than PCA features.
Its architecture was determined through an extensive Bayesian optimization
campaign comprising $120$ trials of a tree-structured Parzen estimator,
searching over network depth, filter counts, kernel sizes, dense layer
dimensions, activation functions, dropout rates, and AdamW optimizer learning
rates. The optimal topology consists of three 1-D convolutional layers ($64$
filters, kernel size $7$, stride $1$, batch normalization, ELU activation, no
spatial pooling) followed by three fully connected layers of $512$ units with ELU
activations. The network is trained with AdamW under a multi-label binary
cross-entropy (BCE) loss function. To explicitly condition the model against
false alarms, a dedicated, disjoint slice of noise-only observations is included
with all-zero target vectors ($\mathbf{y} = \mathbf{0}_{64}$).

\emph{Identical Dataset Benchmark.} Both the SVM and CNN models were
trained, cross-validated, and benchmarked on the identical 1.05-million
sample dataset generated across identical channel conditions. Under five-fold
stratified cross-validation, the CNN achieves $99.04\pm0.02\%$ accuracy and
reaches $99.29\%$ test accuracy (macro-AUC of $1.0000$), matching and slightly
exceeding the SVM offline accuracy. Its $64$ pre-sigmoid logits act directly as
decision scores $\mathbf{s}$, feeding the identical MAD-CFAR thresholding stage.

Table~\ref{tab:svm_vs_cnn} summarizes the head-to-head comparison in the
identical 64-class formulation. In offline classification the CNN is marginally
more accurate -- $0.71$ percentage points -- and markedly cheaper, requiring far
fewer arithmetic operations per occasion and a model over an order of magnitude
smaller, which is attractive for the real-time budget analyzed in
Section~\ref{subsec:complexity}. At the operational, false-alarm-constrained
operating point, however, the two models prove complementary rather than one
strictly dominating: the SVM-based hybrid remains the strongest variant in
missed-detection rate, while the CNN offers comparable behavior in benign channels
at a fraction of the cost (Section~\ref{subsec:cnn_results}). The SVM is retained
as the model on which the explainability analysis (Section~\ref{subsec:xai}) is
carried out, because its kernel-margin structure lends itself to the attribution
and Zadoff--Chu orthogonality study.

\begin{table}[!t]
  \centering
  \caption{Learned detectors in the same 64-class formulation and dataset. The
    CNN is slightly more accurate than the SVM at a fraction of the inference and
    memory cost.
    Cost figures are per PRACH occasion (all $64$ scores).}
  \label{tab:svm_vs_cnn}
  \footnotesize
  \renewcommand{\arraystretch}{1.2}
  \setlength{\tabcolsep}{4pt}
  \resizebox{\columnwidth}{!}{%
  \begin{tabular}{l|cc}
    \hline
    \textbf{Metric} & \textbf{SVM (OvA, poly-2)} & \textbf{CNN (1-D)}\\\hline
    Test accuracy         & $98.58\%$          & $99.29\%$\\
    Macro-averaged AUC    & $0.9991$           & $1.0000$\\
    5-fold CV accuracy    & $98.42\pm0.03\%$   & $99.04\pm0.02\%$\\
    Inference cost        & $\approx1158$~MFLOPs & $\approx154$~MFLOPs\\
    Model size (float32)  & $\approx2.3$~GB    & $\approx112$~MB\\\hline
  \end{tabular}%
  }
\end{table}

\subsection{Explainability}\label{subsec:xai}
To verify \emph{which} signal features drive the decisions, post-hoc
explainability methods are applied. A gradient-sensitivity analysis proved
uninformative for the degree-2 polynomial kernel, whose gradient is not exposed
in the original feature space but in the reproducing-kernel Hilbert space; the
model-agnostic SHAP method~\cite{lundberg2017} is therefore used for global
attribution, with LIME~\cite{ribeiro2016lime} providing complementary local
explanations.
Projecting the PCA-domain importances back onto the real/imaginary signal
samples (a linear map) shows that importance is spread across the entire
sequence, with almost identical mass on the real and imaginary parts
($<0.07\%$ difference) and no dominant time window -- consistent with the
constant-amplitude, zero-autocorrelation (CAZAC) property of the ZC sequences
and indicating that the classifier exploits the global signal structure rather
than a few coordinates.

\begin{figure}[!t]
  \centering
  \includegraphics[width=\columnwidth]{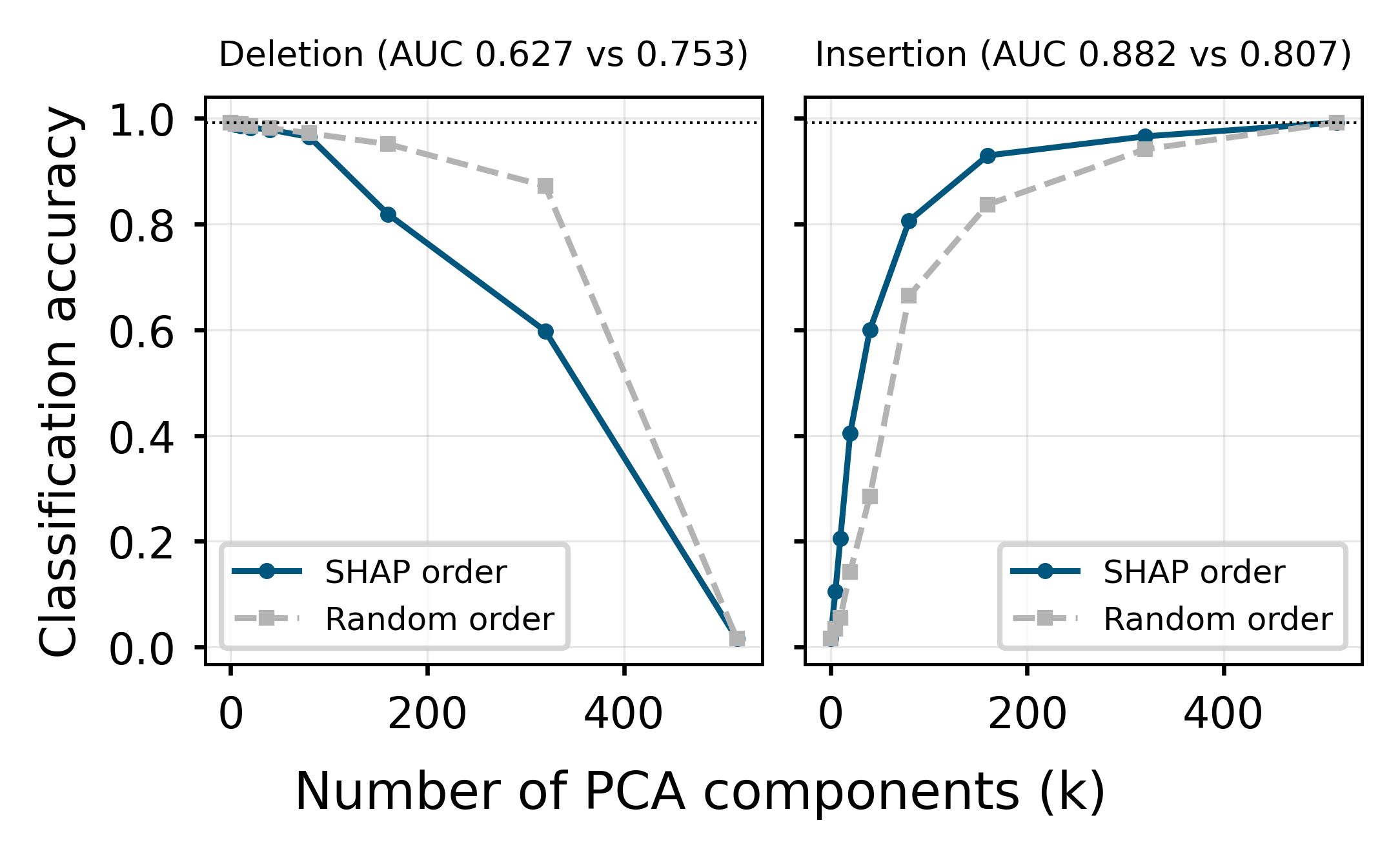}
  \caption{Faithfulness test: accuracy as a function of the number of deleted
    (left) or inserted (right) PCA components, ordered by SHAP importance
    (colored) or at random (gray). The gap between the SHAP and random curves
    confirms that the attributions are faithful to the model.}
  \label{fig:faithfulness}
\end{figure}

\textit{Faithfulness.} To confirm that the SHAP ranking identifies genuinely
influential features rather than method artifacts, an
insertion--deletion faithfulness test~\cite{petsiuk2018rise} is applied on $8\,000$
stratified test samples (baseline accuracy $99.2\%$), comparing the SHAP feature
order against a random-order baseline (Table~\ref{tab:faithfulness},
Fig.~\ref{fig:faithfulness}). Deleting the top-$160$ SHAP components drops the
accuracy by $17.3$ percentage points (to $81.9\%$) versus only $4.1$ points (to
$95.1\%$) for random deletion, and inserting the top-$40$ SHAP components already
recovers $59.9\%$ accuracy versus $28.4\%$ for random insertion. The
insertion/deletion AUCs ($0.882$ vs.\ $0.807$ and $0.627$ vs.\ $0.753$, where a
lower deletion AUC is better) quantitatively confirm that the SHAP attributions
are faithful to the model.

\begin{table}[!t]
  \centering
  \caption{Faithfulness test: classification accuracy versus the number $k$ of
    modified PCA components. SHAP rows ablate in decreasing-importance order;
    \emph{rand.} rows use a random order.}
  \label{tab:faithfulness}
  \footnotesize
  \setlength{\tabcolsep}{3.5pt}
  \resizebox{\columnwidth}{!}{%
  \begin{tabular}{l|ccccccc}
    \hline
    $k$ & 0 & 20 & 40 & 80 & 160 & 320 & 515\\\hline
    Del.\ SHAP  & 0.992 & 0.983 & 0.979 & 0.965 & \textbf{0.819} & \textbf{0.598} & 0.016\\
    Del.\ rand. & 0.992 & 0.986 & 0.982 & 0.972 & 0.951 & 0.872 & 0.016\\\hline
    Ins.\ SHAP  & 0.016 & \textbf{0.404} & \textbf{0.599} & \textbf{0.806} & 0.930 & 0.966 & 0.992\\
    Ins.\ rand. & 0.016 & 0.142 & 0.284 & 0.664 & 0.837 & 0.942 & 0.992\\\hline
  \end{tabular}%
  }
\end{table}

\textit{Zadoff--Chu orthogonality.} An OvA SHAP structure analysis flagged
preambles $14$ and $15$ as the strongest off-diagonal coupling. To test whether
this reflects a genuine similarity of the underlying ZC sequences or is merely a
classifier artifact, their analytical cyclic cross-correlation was computed for
$N_{\mathrm{ZC}}=839$. The measured peak,
$|R_{14,15}|_{\max}/N=0.0345$ ($-29.24$~dB), equals the theoretical value
$1/\sqrt{N_{\mathrm{ZC}}}=0.0345$ that holds for \emph{any} distinct pair of
prime-length ZC preambles. The two sequences are therefore orthogonal in the
cyclic-correlation sense, refuting the hypothesis of a physical similarity: the
off-diagonal SHAP coupling is a property of the trained OvA decision functions,
not of the ZC sequences themselves.

\subsection{Computational Complexity}\label{subsec:complexity}
\begin{figure}[!t]
  \centering
  \includegraphics[width=0.92\columnwidth]{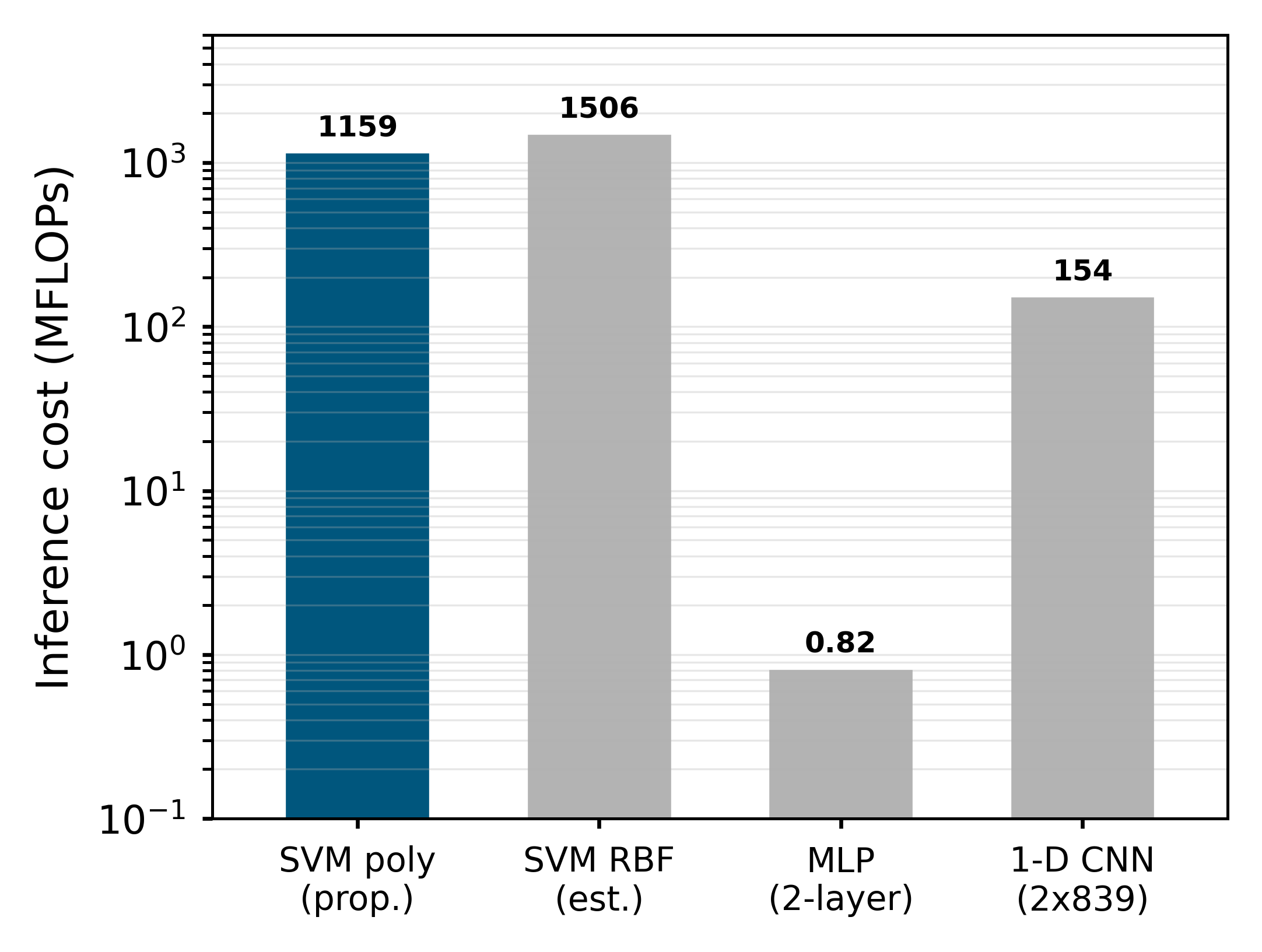}
  \caption{Inference cost (MFLOPs per prediction, log scale) of the proposed
    polynomial-kernel SVM against an RBF-kernel SVM (estimate), a two-layer MLP,
    and a three-layer 1-D CNN. The kernel evaluation dominates the SVM cost.}
  \label{fig:flops}
\end{figure}

Deployment in a real-time BS receiver requires the per-occasion inference cost
to fit the gNB processing budget. The inference pipeline has three stages: PCA
projection, One-vs-All kernel evaluation, and argmax. Counting one fused
multiply--add as $2$~FLOPs, the PCA projection from $D=1678$ to $K=515$ costs
$\mathrm{FLOPs}_{\mathrm{PCA}}\approx 2DK\approx 1.73$~MFLOPs, whereas the
degree-2 polynomial kernel evaluated over all support vectors of the $64$ OvA
classifiers costs
$\mathrm{FLOPs}_{\mathrm{kernel}}=N_{\mathrm{cls}}\,\bar{N}_{SV}\,(2K+5)
\approx 1.16$~GFLOPs~\cite{hennessy2017computer,scholkopf2002learning}. The total
is $\approx 1158$~MFLOPs per prediction, of which the kernel evaluation accounts
for $99.7\%$, PCA for $0.15\%$, and argmax for under $0.001\%$. With an average
of $\bar{N}_{SV}\approx 17\,472$ support vectors per binary classifier
($>1.1\times 10^{6}$ in total), the per-class cost is nearly uniform
(coefficient of variation $\approx 0.93\%$), so pruning low-cost classes would
not meaningfully reduce the budget. The model stores its support vectors in
$\approx 4.6$~GB at float64, halved to $\approx 2.3$~GB at float32 without
accuracy loss; both figures preclude deployment on memory-constrained
microcontrollers or low-end FPGAs without dedicated compression.
Fig.~\ref{fig:flops} compares the proposed polynomial-kernel SVM with
alternative classifiers: it is slightly cheaper than an RBF-kernel SVM
($\approx 1506$~MFLOPs) but far more expensive than compact neural models. The
CNN adopted in Section~\ref{subsec:training} (three convolutional and three dense
layers) costs $\approx154$~MFLOPs per occasion ($\approx77$~M multiply--accumulates:
$97.7$~MFLOPs in the convolutional layers and $56.1$~MFLOPs in the dense layers) --
roughly $8\times$ below the SVM -- and stores its $\approx28$~million
parameters in only $\approx112$~MB, over an order of magnitude smaller than the
SVM's $2.3$~GB of support vectors. Even lighter designs (a two-layer MLP at
$\approx0.82$~MFLOPs, or a pooled CNN at $\approx10$~MFLOPs) trade accuracy for
cost. The dominant SVM kernel cost, by contrast, identifies support-vector
reduction as its main deployment obstacle and is a strong argument for the CNN
as the low-cost branch.

Beyond the operation count, the end-to-end inference latency of a production
C++/cuBLAS SVM implementation was measured on the same H100. For the target
workload of one PRACH occasion ($B=64$ preambles) the median SVM latency is
$2.81$~ms with the model resident in device memory -- about $39\%$ of the
roofline peak and $\sim\!59\times$ faster than a $16$-thread CPU baseline. This
exceeds the indicative $1$~ms gNB detection budget by roughly $2.8\times$, so the
\emph{SVM} does not by itself satisfy a hard real-time constraint even on a
server-class accelerator. The CNN, measured on the same H100 in fp32, processes a
single PRACH occasion in a median of $258$~\textmu s ($257$~\textmu s median,
$267$~\textmu s at the $99$th percentile) -- about $3.9\times$ inside the $1$~ms
budget and roughly $10.9\times$ faster than the SVM; batched over $4096$ occasions
it drops to $4.55$~\textmu s per occasion ($\sim\!220{,}000$ occasions per
second). It therefore comfortably meets the timing budget, which makes it the more
attractive branch from a pure cost standpoint. In either case
the learned model is a \emph{parallel auxiliary} branch, whereas the low-cost
analytical detector alone already meets the timing budget: when the classifier
cannot complete within the window, a deadline-aware fusion emits the analytical
decision only, trading a few decibels of low-SNR sensitivity for guaranteed
timeliness.

\subsection{Adaptive Score-Domain Thresholding: MAD-CFAR}\label{subsec:madcfar}
The core methodological contribution in the score domain is a robust adaptive
threshold applied to the vector
$\mathbf{s}=[s_1,\dots,s_{64}]^{\top}$ of per-class classifier scores of a single
PRACH occasion; these are SVM decision values or, equivalently, CNN logits, so
the mechanism is agnostic to the underlying model. Because the $k\cdot\mathrm{MAD}$
term is dimensionless, only the two parameters $(k,s_\mathrm{min})$ are
re-calibrated per classifier. Motivated by an exploratory analysis of the SVM
scores, whose true-preamble values have a median around $15$--$20$ while
noise-only values rarely exceed $\sim 5.7$, the \emph{MAD-CFAR} threshold is
defined as
\begin{equation}
  T=\max\!\bigl(\widetilde{s}+k\cdot 1.4826\cdot\mathrm{MAD}(\mathbf{s}),\;
  s_\mathrm{min}\bigr),
  \label{eq:robust_threshold}
\end{equation}
where $\widetilde{s}=\mathrm{median}(\mathbf{s})$ estimates the background level,
$\mathrm{MAD}(\mathbf{s})=\mathrm{median}(|s_i-\widetilde{s}|)$ is the robust
dispersion estimator, and $1.4826$ makes $1.4826\cdot\mathrm{MAD}$ an
asymptotically unbiased estimator of the standard deviation for Gaussian
data~\cite{Hampel1986}. Preamble $i$ is detected iff $s_i>T$. The multiplier
$k$ plays the role of the CFAR coefficient $\alpha$, while the absolute floor
$s_\mathrm{min}$ guards against the case in which many preambles have elevated
scores simultaneously, which would inflate the median. The $64$ scores act as
the cells of a CFAR window, and the median is a robust order-statistic estimator
of the background, in analogy with OS-CFAR~\cite{Rohling1983,Gandhi1988}.

The threshold parameters are calibrated on a dedicated dataset, \emph{separate
from the SVM training set}, comprising a signal campaign ($\sim\!248\,000$
preamble occasions) and a noise-only campaign ($\sim\!28$ million positions with
no preamble transmitted), both spanning all six propagation channels. The pair
$(k,s_\mathrm{min})$ is selected by a grid search over $442$
combinations, using a hierarchical criterion that prioritizes a zero false
alarm rate: among all pairs achieving $\mathrm{FAR}_\mathrm{total}=0$ on the
$\sim\!28$ million noise positions, the one maximizing the detection rate is
chosen. The optimum is $k^{\ast}=5.5$ and $s_\mathrm{min}^{\ast}=11.0$, yielding
a detection rate of $91.23\%$ at pre-calibrated zero false alarms.

\begin{figure}[!t]
  \centering
  \includegraphics[width=0.92\columnwidth]{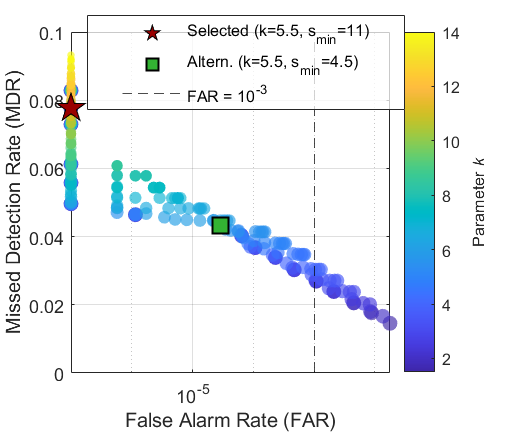}
  \caption{FAR versus MDR trade-off over all tested $(k,s_\mathrm{min})$ pairs.
    The selected operating point ($k=5.5$, $s_\mathrm{min}=11.0$) attains zero
    FAR; the alternative point ($k=5.5$, $s_\mathrm{min}=4.5$) trades a higher
    detection rate for a nonzero FAR. The dashed line marks the
    $\mathrm{FAR}\le 10^{-3}$ requirement.}
  \label{fig:far_vs_mdr}
\end{figure}

\emph{Noise Floor vs. Dynamic Dispersion.} In quiet AWGN/EPA channels where
background score dispersion is small ($\mathrm{MAD}(\mathbf{s}) \approx 1.0$),
the term $\widetilde{s} + k\cdot1.4826\cdot\mathrm{MAD}(\mathbf{s}) \approx 8.15 < 11.0$,
so $s_\mathrm{min}^* = 11.0$ acts as a conservative noise floor protecting the
receiver against false triggers. Under non-homogeneous background conditions,
multi-user access attempts, or elevated interference where background score
dispersion expands ($\mathrm{MAD}(\mathbf{s}) > 1.35$), the dynamic term
$\widetilde{s} + 5.5 \times 1.4826 \times \mathrm{MAD}(\mathbf{s})$ exceeds
$11.0$, elevating $T$ dynamically to maintain constant false alarm probability.
For deployment scenarios permitting a controlled, non-zero false alarm rate
($P_\mathrm{FA} \le 10^{-3}$ per 3GPP requirements), the alternative calibrated
operating point $k=5.5, s_\mathrm{min}=4.5$ elevates detection sensitivity to
$94.85\%$ while keeping $\mathrm{FAR}=9.2\times 10^{-4}$ with continuously active
MAD thresholding (Fig.~\ref{fig:far_vs_mdr}). The identical procedure applied to
CNN logits yields $k^{\ast}=3.0$. Under a like-for-like comparison with the SVM
rule, the CNN records $\mathrm{FAR}=0$ on $128{,}000$ real AWGN-noise positions,
$200{,}000$ synthetic AWGN positions, and a $100{,}000$-position noise campaign
($95\%$ Clopper--Pearson upper bound $\approx3\times10^{-5}$).

\subsection{Hybrid PRACH Receiver}\label{subsec:hybrid}
The two detectors have complementary regions of reliable operation and largely
independent error characteristics, which motivates a fusion rule that lets them
cover each other's weaknesses without any arbitration mechanism. The hybrid
receiver applies a per-preamble logical OR to the binary decisions of the
analytical detector, $d_i^{\mathrm{ana}}$, and the ML detector,
$d_i^{\mathrm{ML}}$:
\begin{equation}
  d_i^{\mathrm{hyb}}=d_i^{\mathrm{ana}}\;\vee\;d_i^{\mathrm{ML}}.
  \label{eq:hybrid_or}
\end{equation}
When both detectors agree, or when only the analytical branch reports a detection, the
analytical detection (with its exact TA) is kept. When only the ML branch
reports a detection, a new detection is added and its TA is estimated \emph{post hoc} from the
same PDP by mapping the preamble index to its (ZC sequence, cyclic shift) pair,
extracting the search window, smoothing it, applying a peak-to-mean quality
gate, locating the first path, and refining it with parabolic sub-sample
interpolation; the result is scaled to $N_\mathrm{TA}$ units per 3GPP
TS~38.213~\cite{3GPP-TS-38.213}. No veto mechanism is used, so every final
decision is directly attributable to one component detector, keeping the
receiver modular and fully interpretable. Algorithm~\ref{alg:hybrid} summarizes
the pipeline.

\begin{algorithm}[!t]
\caption{Hybrid ML--CFAR PRACH detection for one occasion.}
\label{alg:hybrid}
\begin{algorithmic}
\STATE \textbf{Input:} received signal; PDP; classifier scores $\mathbf{s}$ (SVM
       decision values or CNN logits)
\STATE \textbf{Analytical branch:} compute CFAR threshold on the PDP;
       obtain $d_i^{\mathrm{ana}}$ and exact TA
\STATE \textbf{ML branch:} $T\gets\max(\widetilde{s}+k\,1.4826\,\mathrm{MAD}
       (\mathbf{s}),s_\mathrm{min})$;\; $d_i^{\mathrm{ML}}\gets[\,s_i>T\,]$
\FOR{$i=1$ to $64$}
  \STATE $d_i^{\mathrm{hyb}}\gets d_i^{\mathrm{ana}}\vee d_i^{\mathrm{ML}}$
  \IF{$d_i^{\mathrm{ana}}=0$ and $d_i^{\mathrm{ML}}=1$}
     \STATE estimate TA for preamble $i$ post hoc from the PDP
  \ENDIF
\ENDFOR
\STATE \textbf{return} detections $\{d_i^{\mathrm{hyb}}\}$ and TA estimates
\end{algorithmic}
\end{algorithm}

The OR rule detects a preamble whenever either branch does, so the hybrid misses
a preamble only when both branches miss it. Consequently the hybrid missed
detection rate obeys
\begin{equation}
  P_\mathrm{MD}^{\mathrm{hyb}}\le\min\!\bigl(P_\mathrm{MD}^{\mathrm{ana}},
  P_\mathrm{MD}^{\mathrm{ML}}\bigr),
  \label{eq:or_md_bound}
\end{equation}
a guarantee that holds unconditionally, regardless of any statistical dependence
between the two detectors. If the two misses are additionally treated as
independent, the rate factorizes as
\begin{equation}
  P_\mathrm{MD}^{\mathrm{hyb}}=P_\mathrm{MD}^{\mathrm{ana}}\cdot
  P_\mathrm{MD}^{\mathrm{ML}},
  \label{eq:or_md}
\end{equation}
which represents the best case and the source of the expected synergy. When the
branch errors are positively correlated -- as is likely at very low SNR, where
both detectors tend to fail together -- the true hybrid MDR lies between the
product~\eqref{eq:or_md} and the bound~\eqref{eq:or_md_bound}, so
The false alarm rate obeys the union bound
\begin{equation}
  P_\mathrm{FA}^{\mathrm{hyb}}\le P_\mathrm{FA}^{\mathrm{ana}}+
  P_\mathrm{FA}^{\mathrm{ML}},
  \label{eq:or_fa}
\end{equation}
which shows that an un-gated OR receiver cannot achieve a false alarm rate below
the maximum of its components. The construction is practical because the
score-thresholded ML branch has a near-zero false alarm rate ($P_\mathrm{FA}^\mathrm{ML} \approx 0$),
meaning the hybrid FAR effectively coincides with the reference analytical detector's
level across standard scenarios. However, in extreme multi-user or high power-offset
scenarios where the analytical correlation detector experiences false alarm spikes
($P_\mathrm{FA}^\mathrm{ana} > 10^{-3}$), the hard OR rule inherits these spikes. To
eliminate inherited analytical spikes in commercial gNB DUs, generalized
confidence-weighted soft fusion or PDP quality-gated OR rules can be integrated
to suppress marginal analytical triggers when the ML classifier confidently
attributes the occasion to noise.

\section{Experimental Setup}\label{sec:setup}
The evaluation uses the same Nokia LTE/5G physical-layer simulator that
generated the dataset. The receiver variants displayed in
Figs.~\ref{fig:mdr}--\ref{fig:far} are: \textbf{ANALYTICAL} (the reference CFAR
detector on the PDP), \textbf{SVM} (the classifier with the MAD-CFAR threshold of
Section~\ref{subsec:madcfar}), \textbf{CNN} (the lower-cost learned branch under
the same MAD-CFAR rule), and \textbf{HYBRID} (the OR fusion of the analytical and
SVM branches of Section~\ref{subsec:hybrid}). Because the thresholding and fusion
are classifier-agnostic, the same MAD-CFAR pipeline instantiates both learned
branches. Each variant is tested independently over the six
propagation channels across the full SNR range, and additionally on a
conformance-style set of $258$ scenarios derived from an internal PRACH
receiver qualification methodology, covering bandwidth, power-level offset,
frequency offset, multi-antenna, multi-user, and PUSCH co-channel cases. For
each scenario the lowest SNR reaching $P_\mathrm{MD}\le 10^{-2}$
($\mathrm{SNR}_\mathrm{MD}$) and $P_\mathrm{FA}\le 10^{-3}$
($\mathrm{SNR}_\mathrm{FA}$) is recorded, together with the extreme values
$P_\mathrm{MD}^{\min}$ and $P_\mathrm{FA}^{\max}$.
The learned branch is evaluated under two thresholding regimes for the CNN. Its
MAD-CFAR operating point was recalibrated on the six-channel noise so that
$P_\mathrm{FA}\le10^{-3}$ is respected in every channel; the resulting operational
MDR/FAR-vs-SNR curves are reported alongside the SVM and analytical baselines in
Figs.~\ref{fig:mdr}--\ref{fig:far}. Because the SVM is the quality-optimal learned
branch, the full $258$-scenario conformance evaluation is carried out for the
SVM-based variants; the CNN, whose value lies in its much lower cost rather than
in higher detection quality, is characterized on the six-channel MDR/FAR sweep.

\emph{Closed-Set Industrial Simulator Justification.} Evaluation of physical-layer
base station receivers under realistic 3GPP conformance conditions necessitates
industrial-grade simulation platforms. Academic open-source software tools (e.g.,
srsRAN, OpenAirInterface) lack commercial-grade baseband correlation pipelines,
high-fidelity multi-antenna spatial channel emulation, and the automated 258-scenario
conformance test benches required for industrial base station qualification. Testing
within Nokia's proprietary physical-layer simulator provides a rigorous,
commercial-grade evaluation environment that reproduces real gNB baseband processing
and conducted conformance specifications under 3GPP TS 38.141-1~\cite{3GPP-TS-38.141-1},
thereby validating deployment readiness for commercial 5G NR base stations.

\emph{Reproducibility.} Although the Nokia physical-layer simulator, the generated
dataset, and the internal conformance suite are proprietary and cannot be released
publicly, the proposed method is fully specified for independent implementation on
equivalent synthetic or captured datasets. All classifier hyperparameters
(Section~\ref{subsec:training}), 1-D CNN architecture details, the MAD-CFAR
thresholding rule~\eqref{eq:robust_threshold} with calibrated coefficients
$(k, s_\mathrm{min})$, and the OR decision fusion (Algorithm~\ref{alg:hybrid}) are
provided in full.

\section{Results}\label{sec:results}
\subsection{MDR and FAR versus SNR}
\begin{figure}[!t]
  \centering
  \includegraphics[width=\columnwidth]{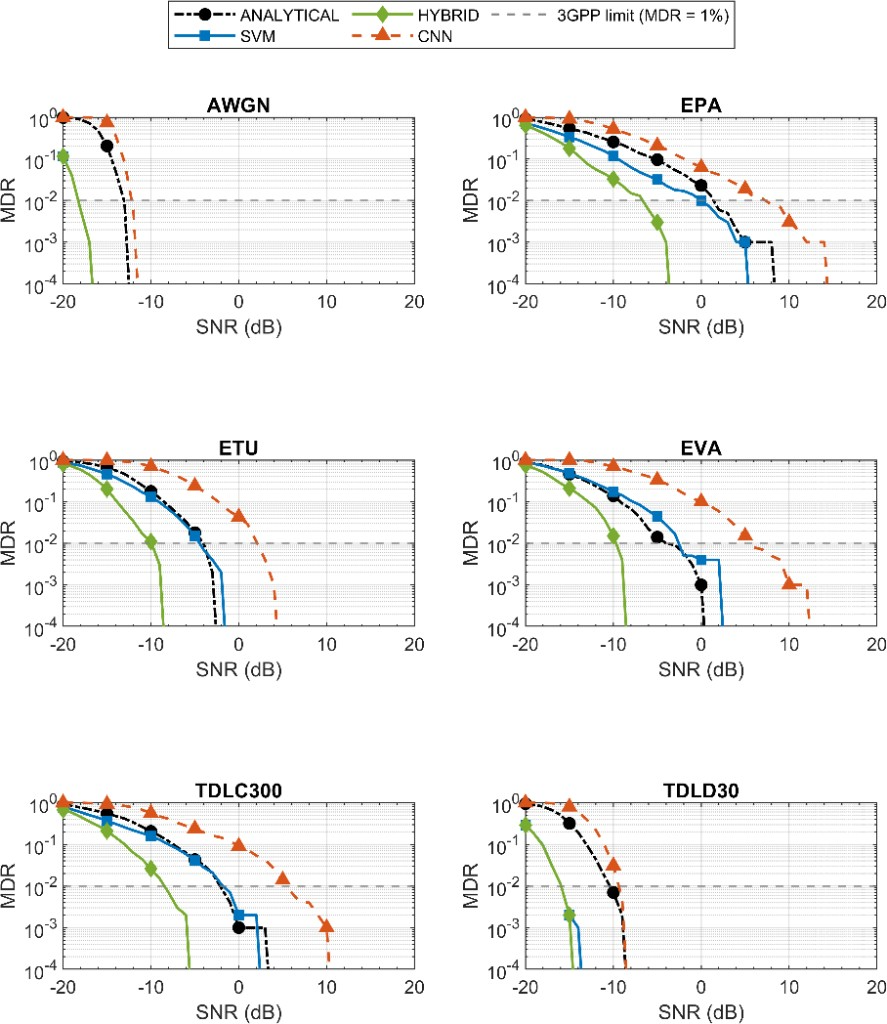}
  \caption{MDR versus SNR for all propagation channels (AWGN, EPA, EVA, ETU,
    TDLC300, TDLD30) and all receiver variants (ANALYTICAL, SVM, CNN, and the
    SVM-based HYBRID).}
  \label{fig:mdr}
\end{figure}

\begin{figure}[!t]
  \centering
  \includegraphics[width=\columnwidth]{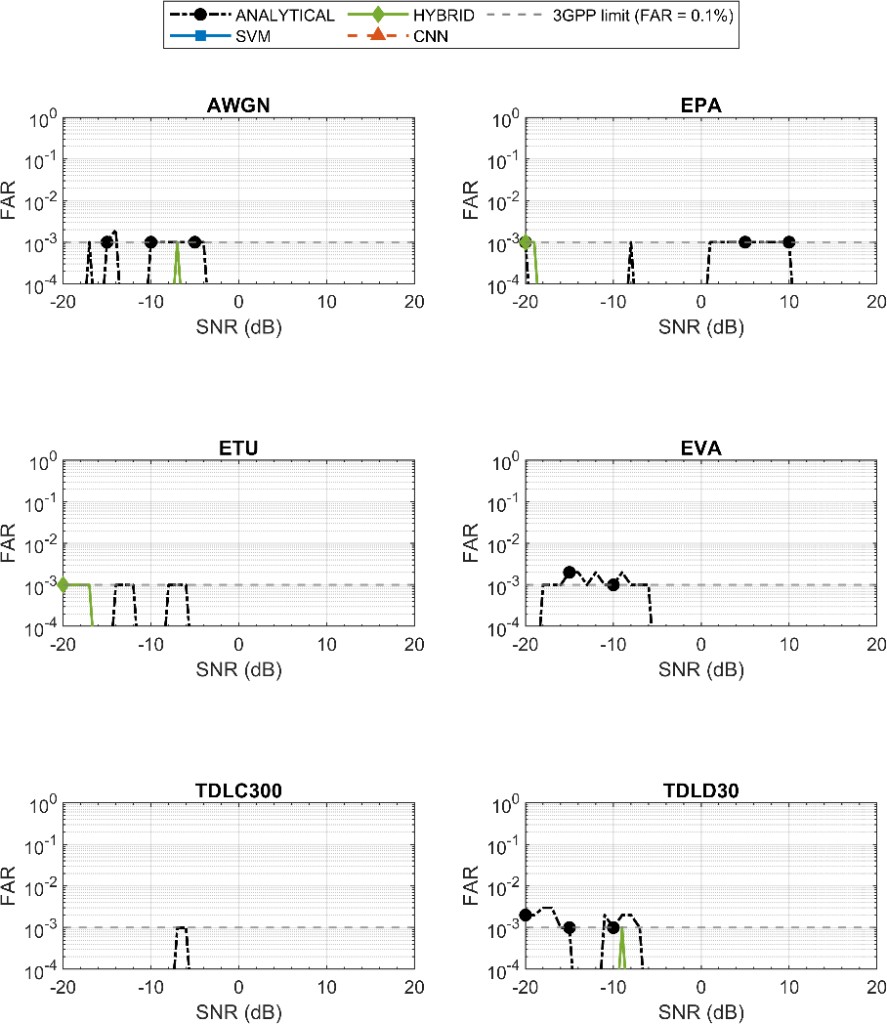}
  \caption{FAR versus SNR for all propagation channels and all receiver variants
    (ANALYTICAL, SVM, CNN, and the SVM-based HYBRID). With the recalibrated
    MAD-CFAR operating point the CNN keeps $\mathrm{FAR}<10^{-3}$ in every
    channel, so its curve stays at the plot floor.}
  \label{fig:far}
\end{figure}

Figs.~\ref{fig:mdr} and~\ref{fig:far} show the MDR and FAR characteristics across
the six channels for all receiver variants: the ANALYTICAL detector, the SVM, the
CNN, and the SVM-based HYBRID (analytical OR SVM). The SVM-based variants and the
hybrid are analyzed below; the CNN behavior is discussed in
Section~\ref{subsec:cnn_results}.

\textit{False alarm control.} The analytical detector keeps
$\mathrm{FAR}<10^{-3}$ over the whole SNR range, as expected from its built-in
CFAR. The improved SVM branch produces no false alarms across the SNR sweep,
consistent with the zero-false-alarm operating point selected on the calibration
campaign (Section~\ref{subsec:madcfar}). Since the OR rule
obeys~\eqref{eq:or_fa}, the hybrid FAR cannot drop below the analytical level;
its FAR is therefore determined by, and effectively equal to, the analytical
detector's low but \emph{nonzero} rate rather than being zero. By contrast, the
legacy SVM without dedicated thresholding reaches a FAR of tens of percent at
low SNR (e.g., $\mathrm{FAR}\approx 0.68$ for ETU at $\mathrm{SNR}=-20$~dB),
which alone disqualifies it under 3GPP requirements.

\textit{Sensitivity.} In the AWGN channel the SVM reaches zero MDR already at
$\mathrm{SNR}=-16$~dB, whereas the analytical detector requires
$\mathrm{SNR}\ge-12$~dB -- a $4$~dB gain. In channels with significant delay
spread (EPA, EVA, ETU, TDLC300) the two detectors are partially complementary,
and their OR fusion yields a clear advantage. For EPA at $\mathrm{SNR}=-20$~dB
the measured MDRs are $0.910$ (analytical), $0.709$ (SVM), and $0.644$
(hybrid), closely matching the value
$P_\mathrm{MD}^{\mathrm{ana}}\!\cdot\!P_\mathrm{MD}^{\mathrm{ML}}=0.645$ implied
by~\eqref{eq:or_md}. A practically important effect is the shift of the
point at which the MDR becomes negligible: in ETU it moves from about $+1$~dB
for the individual detectors to about $-8$~dB for the hybrid, a gain of several
decibels at an unchanged FAR. In TDLD30, dominated by a strong line-of-sight
component, the SVM is markedly better (zero MDR at $-14$~dB versus $-8$~dB for
the analytical detector), and the hybrid characteristic nearly coincides
with the SVM one, again consistent with~\eqref{eq:or_md}.

\subsection{Conformance-Style Stress Tests}
Table~\ref{tab:verdicts} aggregates the $258$ scenarios by the type of outcome,
assigned automatically from $\mathrm{SNR}_\mathrm{MD}$. Because the OR rule
declares a preamble whenever \emph{either} branch does,
$P_\mathrm{MD}^{\mathrm{hyb}}\le\min(P_\mathrm{MD}^{\mathrm{ana}},
P_\mathrm{MD}^{\mathrm{ML}})$ holds by construction; the empirical finding that
the \texttt{HYB\_LOSS} category never occurs therefore confirms that the
implementation respects this guarantee rather than uncovering an unexpected
property. The cost of the OR fusion instead appears on the false-alarm side,
quantified in Table~\ref{tab:selected_consolidated} below.

\begin{table}[!t]
  \centering
  \caption{Aggregated verdicts over the $258$ conformance-style scenarios;
    $N$ is the approximate scenario count per category (exact to within a few
    boundary cases).}
  \label{tab:verdicts}
  \footnotesize
  \renewcommand{\arraystretch}{1.2}
  \setlength{\tabcolsep}{4pt}
  \begin{tabularx}{\columnwidth}{|>{\ttfamily}l|c|X|}
    \hline
    \normalfont\textbf{Verdict} & \textbf{$N$} & \textbf{Interpretation}\\\hline
    SVM\_BEST & $\sim\!110$ & SVM and HYB jointly best; HYB mirrors SVM\\\hline
    ANA\_BEST & $\sim\!25$  & FAR-only and ties; HYB mirrors ANA\\\hline
    ANA\_FAIL & $\sim\!45$  & ANA below threshold; HYB recovers via SVM\\\hline
    SVM\_FAIL & $\sim\!30$  & SVM below threshold; HYB recovers via ANA\\\hline
    ALL\_FAIL & $\sim\!45$  & both below threshold; MDR of order $0.1$--$1.0$\\\hline
    \makecell[l]{HYB\_BEST,\\ HYB\_TIE} & $\sim\!3$ &
      explicit OR synergy (HYB better than either component)\\\hline
    HYB\_LOSS & \textbf{0} & \textbf{HYB never degrades the ANA result}\\\hline
  \end{tabularx}
\end{table}

Table~\ref{tab:selected_consolidated} aggregates representative scenarios
across the 258-scenario conformance test suite, presenting missed detection
($\mathrm{SNR}_\mathrm{MD}$, $P_\mathrm{MD}^{\min}$) and false alarm
($\mathrm{SNR}_\mathrm{FA}$, $P_\mathrm{FA}^{\max}$) metrics side-by-side.
The hybrid receiver consistently matches the better component and, in the
frequency-offset case, strictly outperforms both ($\mathrm{SNR}_\mathrm{MD}$
improved by $1$~dB), demonstrating OR decision gains. When the SVM fails
(e.g., 4-RX PUSCH co-channel case due to unmodeled interference), the
hybrid falls back to the analytical detector; when the analytical detector fails
(e.g., bandwidth, power-offset, and multi-user cases), the hybrid recovers
detection through the SVM. The hybrid keeps $P_\mathrm{FA}^{\max}\le 10^{-3}$ in
most cases, though in line with the union bound~\eqref{eq:or_fa}, it inherits the
analytical detector's occasional FA spikes in extreme scenarios. Across the full
set, $P_\mathrm{FA}\le 10^{-3}$ is satisfied in approximately $225$ of the $258$
scenarios.

\begin{table*}[t]
  \centering
  \caption{Performance Comparison across Representative Conformance Scenarios. A dash (--) indicates that the detector failed to reach $P_\mathrm{MD}\le 10^{-2}$ or $P_\mathrm{FA}\le 10^{-3}$.}
  \label{tab:selected_consolidated}
  \footnotesize
  \renewcommand{\arraystretch}{1.25}
  \setlength{\tabcolsep}{5pt}
  \resizebox{\textwidth}{!}{%
  \begin{tabular}{l|ccc|ccc|ccc|ccc}
    \hline
    \multirow{2}{*}{\textbf{Scenario}} &
    \multicolumn{3}{c|}{$\mathbf{SNR}_\mathbf{MD}$ \textbf{[dB]}} &
    \multicolumn{3}{c|}{$\mathbf{P}_\mathbf{MD}^{\mathbf{\min}}$} &
    \multicolumn{3}{c|}{$\mathbf{SNR}_\mathbf{FA}$ \textbf{[dB]}} &
    \multicolumn{3}{c}{$\mathbf{P}_\mathbf{FA}^{\mathbf{\max}}$} \\
    \cline{2-13}
    & \textbf{ANA} & \textbf{SVM} & \textbf{HYB} & \textbf{ANA} & \textbf{SVM} & \textbf{HYB} & \textbf{ANA} & \textbf{SVM} & \textbf{HYB} & \textbf{ANA} & \textbf{SVM} & \textbf{HYB} \\
    \hline
    Bandwidth & -- & $-17.0$ & $-17.0$ & 0.035 & 0.000 & 0.000 & $-17.0$ & $-17.0$ & $-17.0$ & 0.000 & 0.000 & 0.000 \\
    Power offset & -- & $-14.1$ & $-14.1$ & 0.015 & 0.000 & 0.000 & $-19.0$ & $-19.0$ & $-19.0$ & 0.005 & 0.000 & 0.005 \\
    Multi-antenna & $-16.0$ & $-17.0$ & $-17.0$ & 0.010 & 0.000 & 0.000 & $-17.0$ & $-17.0$ & $-17.0$ & 0.000 & 0.000 & 0.000 \\
    Freq. offset & $-13.1$ & $-13.2$ & $-14.0$ & 0.000 & 0.000 & 0.000 & $-11.2$ & $-14.0$ & $-11.2$ & 0.005 & 0.000 & 0.005 \\
    PUSCH co-ch. 2RX & -- & $-16.5$ & $-16.5$ & 0.046 & 0.000 & 0.000 & $-16.5$ & $-16.5$ & $-16.5$ & 0.000 & 0.000 & 0.000 \\
    Freq. offset FA & $-14.3$ & -- & $-14.3$ & 0.000 & 0.011 & 0.000 & $-12.0$ & $-17.0$ & $-12.0$ & 0.002 & 0.001 & 0.003 \\
    PUSCH co-ch. 4RX & $-14.1$ & -- & $-14.1$ & 0.008 & 0.394 & 0.008 & $-15.0$ & $-15.0$ & $-15.0$ & 0.002 & 0.000 & 0.002 \\
    Multi-user & -- & $-17.0$ & $-17.0$ & 0.025 & 0.005 & 0.005 & -- & -- & -- & 0.005 & 0.005 & 0.010 \\
    2nd PRACH hi-pwr & -- & -- & -- & 0.088 & 1.000 & 0.088 & $-11.0$ & $-11.0$ & $-11.0$ & 0.000 & 0.000 & 0.000 \\
    \hline
  \end{tabular}%
  }
\end{table*}

\subsection{Comparison with the Legacy Classifier}
Introducing the MAD-CFAR threshold lowers the MDR relative to the legacy
classifier in every channel (e.g., in EPA at $\mathrm{SNR}=-20$~dB from $0.709$
to $0.574$). In addition, it eliminates false alarms entirely: in the
legacy design every misclassification was simultaneously a false alarm and a
missed detection, so its MDR and FAR curves coincided over the whole SNR range.
By rejecting decisions that correspond to noise realizations,
\eqref{eq:robust_threshold} drives the FAR to zero and makes the improved
classifier usable, for the first time, as a component of a BS receiver.

\subsection{CNN Branch: Cost, Accuracy, and Operational Behavior}\label{subsec:cnn_results}
At the classification level the CNN matches and slightly exceeds the SVM in the
identical 64-class formulation ($99.29\%$ vs.\ $98.58\%$ test accuracy;
Table~\ref{tab:svm_vs_cnn}), while requiring roughly an order of magnitude fewer
operations and a model over an order of magnitude smaller. Because the MAD-CFAR
threshold and the OR fusion operate on the classifier scores alone, the entire
receiver pipeline transfers to the CNN without modification, so a CNN-based hybrid
would inherit the same no-regression guarantee~\eqref{eq:or_md_bound} and the same
union-bounded FAR~\eqref{eq:or_fa} as the SVM-based hybrid.

The offline gains do not, however, transfer one-to-one to the operational,
false-alarm-constrained sweep. With its MAD-CFAR operating point recalibrated on
the six-channel noise, the CNN keeps $\mathrm{FAR}<10^{-3}$ in every channel
(Fig.~\ref{fig:far}), matching the tight false-alarm control of the SVM branch. In
missed-detection rate (Fig.~\ref{fig:mdr}) it is competitive with the analytical
detector in AWGN and in the held-out TDLC300 channel, but it currently trails the
SVM by several decibels in the high-delay-spread and high-Doppler channels (EPA,
EVA, ETU, TDLD30). In other words, despite its higher offline accuracy, the CNN's
sensitivity at the FAR-constrained operating point does not yet match the SVM in
the most demanding multipath conditions, and the SVM-based hybrid remains the
strongest variant overall.

We report this as an honest, interim result rather than a finished claim of CNN
superiority. The gap is plausibly attributable to the fixed inference-domain
preprocessing and to the absence of timing-offset and carrier-frequency-offset
augmentation during CNN training, both of which the SVM pipeline was tuned around.
It is expected to narrow with a broader architecture search, inference-domain-matched
preprocessing, recalibration of $(k,s_\mathrm{min})$ on a larger multi-channel noise
campaign, and additional training (Section~\ref{sec:conclusion}). The key takeaway
is that the MAD-CFAR threshold and the OR hybrid are genuinely classifier-agnostic:
they impose FAR control on the CNN exactly as on the SVM, so a lower-cost learned
branch is already viable and its detection sensitivity is an optimization problem,
not a structural limitation of the method.

\subsection{Comparison with Existing Machine Learning Receivers}\label{subsec:lit_comparison}
To contextualize the proposed \emph{SVM + CFAR + OR} receiver within the
broader literature, we examine prior machine-learning-based PRACH detectors.
Although the reviewed literature addresses machine learning for PRACH,
qualitative and conceptual differences separate these works from the current
multi-class hybrid approach.

\textit{Modina et al. (2019)}~\cite{Modina2019} conducted an early study
incorporating machine learning into 5G PRACH reception. Instead of processing raw
IQ samples, their algorithm operates on features extracted from the correlator
output (Power Delay Profile). The ML stage ($k$-NN, Naive Bayes) evaluates a
binary decision (``Is this correlation peak genuine?'') after correlation.
In that design, machine learning acts as an auxiliary post-processor rather than a
primary multi-class detector, and a full SVM implementation was not realized.

\textit{Zehra et al. (Scientific Reports, 2022)}~\cite{Zehra2022} evaluated
classical classifiers (Decision Trees, Naive Bayes, $k$-NN, Bagged Trees) to
distinguish valid correlation peaks from noise at low SNR. However, the
transmitted preamble is assumed known a priori, reducing the task to binary
classification ($k=2$). Conceptually, the difference lies between answering
``Is this a valid peak?'' versus ``Which of the 64 standardized preambles was
transmitted?''.

\textit{Guel et al. (2024)}~\cite{Guel2024} proposed a CNN-based detector
operating on Complex Delay Profiles and correlation maps, primarily aiming to
counter inter-cell interference rather than maximizing receiver sensitivity.
Their CNN solves a binary detection task ($k=2$) rather than full preamble
identification across all 64 classes.

\textit{Singh et al. (2024)}~\cite{Singh2024} proposed a hybrid receiver
combining correlation with a Deep Neural Network (DNN) for timing advance
estimation. Machine learning assists the classical detector in peak validation and
timing estimation rather than replacing preamble detection.

Table~\ref{tab:metaComparison} summarizes the qualitative comparison between
existing approaches and the proposed \emph{SVM + CFAR + OR} receiver.

\begin{table*}[t]
\centering
\caption{Qualitative Comparison of Representative Machine-Learning-Based PRACH Receivers}
\label{tab:metaComparison}
\renewcommand{\arraystretch}{1.25}
\resizebox{\textwidth}{!}{%
\begin{tabular}{lcccccc}
\toprule
\textbf{Work} & \textbf{ML Input} & \textbf{ML Output} & \textbf{FAR Control} & \textbf{Reported Gain} & \textbf{Receiver Role} & \textbf{Direct Comparison} \\
\midrule
Modina et al.~\cite{Modina2019} & Power Delay Profile & Peak validity ($k=2$) & No explicit CFAR & Improved over threshold & Post-processing & Incompatible \\
Zehra et al.~\cite{Zehra2022} & Peak features & True / False peak ($k=2$) & No explicit CFAR & Low-SNR detection gain & Peak classification & Incompatible \\
Guel et al.~\cite{Guel2024} & Complex Delay Profile & Interference ($k=2$) & Empirical threshold & Inter-cell interference & Interference filter & Incompatible \\
Singh et al.~\cite{Singh2024} & Correlator output & Detection / TA ($k=2$) & Partial & Improved robustness & Receiver assistance & Incompatible \\
\textbf{This work} & \textbf{Raw IQ / PCA samples} & \textbf{64 Preambles ($k=64$)} & \textbf{MAD-CFAR} & \textbf{Several-dB MDR gain} & \textbf{Primary Detector} & \textbf{Reference} \\
\bottomrule
\end{tabular}%
}
\end{table*}

\subsection{Scientific Invalidity of Direct Quantitative Comparison \& Meta-Analysis}\label{subsec:invalid_comparison}
An important question is whether the proposed receiver can be quantitatively
compared with previously published ML PRACH detectors in terms of Missed Detection
Rate (MDR) and False Alarm Rate (FAR) numerical values.

The answer is that such direct numerical comparison is \emph{not scientifically
valid}. Although all reviewed publications address PRACH reception, they
formulate different learning tasks, adopt non-identical receiver reference
architectures, and evaluate performance under conflicting experimental assumptions.
A meaningful numerical comparison of MDR and FAR requires that all receivers
evaluate: (i) identical channel models and Doppler spreads, (ii) identical SNR
ranges and power offsets, (iii) identical PRACH preamble formats (Short vs.\
Long), (iv) identical definitions of Missed Detection and False Alarm events, and
(v) identical underlying evaluation datasets. None of the reviewed publications
satisfy these combined conditions.

\textit{Comparison of False Alarm Control.} While numerical FAR values cannot be
compared directly, false alarm control mechanisms differ fundamentally. As shown
in Table~\ref{tab:qualitative_matrix}, prior ML PRACH receivers rely on fixed
classifier decision thresholds or confidence scores without explicit CFAR control.
The proposed receiver is unique in introducing the score-domain \emph{MAD-CFAR}
mechanism, providing analytical and calibrated operational FAR control directly
in score space.

\begin{table}[!t]
\centering
\caption{Qualitative Architectural Feature Matrix of ML PRACH Receivers}
\label{tab:qualitative_matrix}
\renewcommand{\arraystretch}{1.2}
\resizebox{\columnwidth}{!}{%
\begin{tabular}{lcccc}
\toprule
\textbf{Feature} & \textbf{Modina} & \textbf{Zehra} & \textbf{Singh} & \makecell{\textbf{Proposed}\\\textbf{(SVM+CFAR+OR)}} \\
\midrule
Peak detection & \checkmark & \checkmark & \checkmark & \checkmark \\
Preamble identification & -- & -- & -- & \checkmark \\
Explicit score CFAR control & Partial & Partial & Partial & \checkmark \\
Hybrid logical-OR decision & -- & -- & Partial & \checkmark \\
No-regression guarantee & -- & -- & -- & \checkmark \\
\bottomrule
\end{tabular}%
}
\end{table}

\textit{Toward a Meta-Comparison Framework.} Future surveys on ML-based physical
layer receivers should adopt a structured meta-comparison framework rather than
comparing isolated numerical metrics. Recommended criteria include: (i) receiver
architecture integration point, (ii) input signal representation (raw IQ vs.\ PDP
correlation), (iii) learning problem formulation (binary peak validation vs.\
64-class preamble classification), (iv) score-domain false-alarm control mechanism,
(v) computational complexity per occasion, and (vi) 3GPP TS 38.141-1 conducted
conformance test readiness. Such qualitative meta-analysis accurately reflects
scientific progress and positions learned receiver architectures within 5G NR and
6G standards.

\section{Conclusion}\label{sec:conclusion}
This article presented the \emph{SVM + CFAR + OR} hybrid PRACH receiver and
established a broader, classifier-agnostic \emph{ML + CFAR + OR} framework
built around two key mechanisms: a robust adaptive MAD-CFAR threshold operating in
the domain of raw classifier decision scores, and a logical-OR decision fusion with
a classical PDP-CFAR detector. The MAD-CFAR mechanism drives the learned branch's
false alarm rate to zero on a 28-million-position noise calibration campaign, while
the OR hybrid inherits the analytical detector's standard-compliant false alarm rate
while extending sensitivity by several decibels. Across six propagation channels
and 258 conformance scenarios evaluated in Nokia's commercial physical-layer
simulator, the \emph{SVM + CFAR + OR} receiver exhibited zero missed-detection
regression relative to the reference detector -- an analytical property
guaranteed by the OR construction. Model independence within the
\emph{ML + CFAR + OR} framework was demonstrated by evaluating a lightweight 1-D
CNN on the identical dataset, achieving $99.29\%$ offline accuracy at an
order-of-magnitude lower compute and memory footprint. These findings confirm that
score-thresholded machine learning models can be safely integrated into the
AI-native radio interfaces envisioned for 6G networks.

\emph{Limitations.} The study also exposes clear limitations. The learned branch
is trained without carrier-frequency-offset realizations and consequently
degrades markedly under residual frequency offset ($P_\mathrm{MD}^{\min}$ of order
$0.4$--$1.0$ in the affected scenarios); the hybrid masks this at the system
level by falling back to the analytical branch, but the classifier itself
remains vulnerable. Because the OR rule cannot lower the false-alarm rate below
the analytical baseline, the hybrid inherits the analytical detector's sporadic
FA spikes and exceeds $P_\mathrm{FA}=10^{-3}$ in a minority of the conformance
scenarios (about $33$ of $258$). On the cost side, the SVM is large
($\approx 4.6$~GB) and its $2.81$~ms latency exceeds the gNB timing budget,
whereas the CNN reduces both by more than an order of magnitude and, at a measured
$258$~\textmu s per occasion on the same H100, meets the budget with margin. The
CNN's own operational sensitivity is a further limitation: at the FAR-constrained
operating point it
currently trails the SVM by several decibels in the high-delay-spread and
high-Doppler channels, so its lower cost does not yet translate into a
best-in-class detector, and the reported CNN result should be read as an interim,
not-yet-optimized baseline. The timing-advance estimate for ML-only detections is
produced by the hybrid's post-hoc PDP re-estimation and its accuracy is not
separately quantified here. Finally, the entire evaluation is simulation-based;
validation on real radio-frequency captures remains outstanding.

Future work includes closing the CNN's operational sensitivity gap through a
broader architecture search, inference-domain-matched preprocessing, and
recalibration of the MAD-CFAR $(k,s_\mathrm{min})$ on a larger multi-channel noise
campaign; augmenting the training set with carrier-frequency-offset, additional
timing-advance, and preamble-format realizations; further reducing the CNN cost
(e.g., through pooling or pruning) to widen the real-time margin; a dedicated
evaluation of the post-hoc timing-advance accuracy; and validation on real
radio-frequency captures.
A further direction is a confidence-aware fusion rule that combines the two
branches by their soft, per-occasion confidence rather than their hard
decisions; generalizing the logical OR in this way would allow the receiver to
suppress marginal analytical detections that the classifier confidently
attributes to noise, and thus push the false-alarm rate below the analytical
baseline at the cost of the strict no-regression guarantee. Online adaptation of
the threshold parameters and hardware-in-the-loop validation on real radio
frequency complete the roadmap.

\section*{Acknowledgment}
The authors thank Nokia for access to the LTE/5G physical-layer simulator.
Created using resources provided by Wroc{\l}aw Centre for Networking and
Supercomputing (\url{http://wcss.pl}).

\end{document}